# Modeling a Century of Citation Distributions

Matthew L. Wallace, Vincent Larivière, Yves Gingras

Observatoire des sciences et des technologies (OST), Centre interuniversitaire de recherche sur la science et la technologie (CIRST), Université du Québec à Montréal, Case Postale 8888, succ. Centre-Ville, Montréal (Québec), H3C 3P8, Canada. E-mail: mattyliam@gmail.com; lariviere.vincent@uqam.ca; gingras.yves@uqam.ca

## Abstract

**Background**

Changes in citation distributions over 100 years can reveal much about the evolution of the scientific communities or disciplines. The prevalence of uncited papers or of highly-cited papers, with respect to the bulk of publications, provides important clues as to the dynamics of scientific research.

**Key Findings**

Using 25 million papers and 600 million references from the Web of Science over the 1900-2006 period, this paper proposes a simple model based on a random selection process to explain the "uncitedness" phenomenon and its decline in recent years. We show that the proportion of uncited papers is a function of 1) the number of articles published in a given year (the competing papers) and 2) the number of articles subsequently published (the citing papers) and the number of references they contain. Using uncitedness as a departure point, we demonstrate the utility of the stretched-exponential function and a form of the Tsallis function to fit *complete* citation distributions over the 20$^{th}$ century. As opposed to simple power-law fits, for instance, both these approaches are shown to be empirically well-grounded and robust enough to better understand citation dynamics at the aggregate level.

**Conclusions**

Based on an expansion of these models, on our new understanding of uncitedness and on our large dataset, we are able provide clear quantitative evidence and provisional explanations for an important shift in citation practices around 1960, unmatched in the 20th century. We also propose a revision of the "citation classic" category as a set of articles which is clearly distinguishable from the rest of the field.

# Introduction

Since Price's pioneering work on networks of citations [1] and his subsequent development of the cumulative advantage model to explain observed power-law behavior [2], much effort has gone into understanding the mathematics needed to characterize the citation distribution [3, 4] and to define its underlying mechanisms [5-7]. However, these distributions have been studied either from a mainly theoretical perspective [5-9] or empirically, using a dataset of references made from a few years of data or based on a small portion of the scientific community [3, 4, 10, 11]. Redner's examination of complete citation distributions of a century of articles published in *Physical Review* and of articles published in 1981 has notably shown that, for a small number of citations, a stretched exponential provides a better fit, while power-law behavior dominates at a high number of citations [3, 4]. Mathematical models have also succeeded in gaining some insight, *ab initio*, into how these distributions arise, but for the most part, these approaches come at the cost of high levels of parameterization and cannot explain the whole range of the citation distribution, especially the stretched-exponential regime [5, 6]. Given that most of these studies used age distribution of *cited* material for a given year, by definition, *uncited* articles were often excluded from the analysis. Many models based on citation network growth also fail to include the zero-citation case [2, 11]. But if one wants to study complete citation distributions, the natural point of departure would be uncited papers which, it is often erroneously believed, constitute the majority of the scientific literature [12-15]. Although several empirical studies [16-21] challenged this belief using data for a few fields and small periods of time, no study has yet measured the changes in scientific articles' citedness and uncitedness rates over a long period of time and across fields.

This paper analyzes the changes in articles' citation rates over a 107-year period (1900-2006) for all fields of the natural sciences and engineering (NSE), medicine (MED), social sciences (SS) and humanities (HUM). The following section briefly presents the methods; section 3 presents empirical measures of articles' evolving "citedness" and uncitedness rates. It also provides empirical measures of the mean and median citation rates of articles. Based on these data, we focus on the specific case of uncitedness and on *complete* models of citation distributions. Given that data for the full period is only available for NSE and MED, our examination of uncitedness will be centered on these two fields, and the bulk of our subsequent analysis (Sections 5-7) will be limited to NSE data. We demonstrate the importance and utility of the stretched-exponential regime of the citation probability distribution function, and use it to collapse 100 years of citation data onto two master curves. We also demonstrate the robustness of a mathematical model derived from a generalization of Boltzmann-Gibbs statistics and propose an alternative definition of the "citation classic". We achieve these goals using simple approximations and models with few adjustable parameters, allowing us to straightforwardly elucidate trends and characteristics that have dominated citation practices since 1900. Based on a wealth of empirical data, this paper makes use of sociological explanations and mathematical modeling — two approaches that are rarely combined in bibliometrics — in order to explain the changes in articles' citation rates over the last century.

## Methods

Data for this paper are drawn from Thomson Scientific's Web of Science, including the Century of Science database, along with the Science Citation Index Expanded (SCIE), Social Sciences Citation Index (SSCI) and Arts and Humanities Citation Index (AHCI). To our knowledge, this is by far the most comprehensive and systematic study of citation statistics. For NSE and MED, it covers the period 1900-2006, while the data for the social sciences and for the humanities start in 1956 and 1975, respectively. Only citations received by research articles, notes and review articles are included in the study; self-citations were also excluded. Citations made to 12 million (M) papers in NSE, 10 M in MED, 2 M in SS and 1 M in HUM are retrieved in a pool of 600 M references recorded in the database using 2-, 5- and 10-year citation windows. Journal classification is based on that used by the National Science Foundation, complemented with some in-house classification for the humanities. The matching of citations to source articles was made using Thomson's reference identifier provided with the data, as well as additional matching using the author, publication year, volume number and page numbers.

## Analysis

**Trends in citation distributions**

Fig. 1 presents the variations, over the period studied, of the mean number of citations received by papers for both citation windows. The median values, less affected by the highly cited papers, follow the same trends. One can readily notice that, since the end of the sixties, there has been a striking increase in the number of citations received by papers, except in HUM where the numbers are stable. MED is the field in which papers are most cited, followed by NSE, SS and HUM. An interesting and apparent feature of MED and NSE curves is a decrease in the mean and median number of citations received around World Wars I and II, due to a rapid decline in the number of articles published [22, 23]. But perhaps more interesting is that, in these two fields, the post-WWII increase in articles' citation rates was followed by a 5- or 10-year decrease around 1960. This surprising phenomenon will be discussed in Section 7 of this paper.

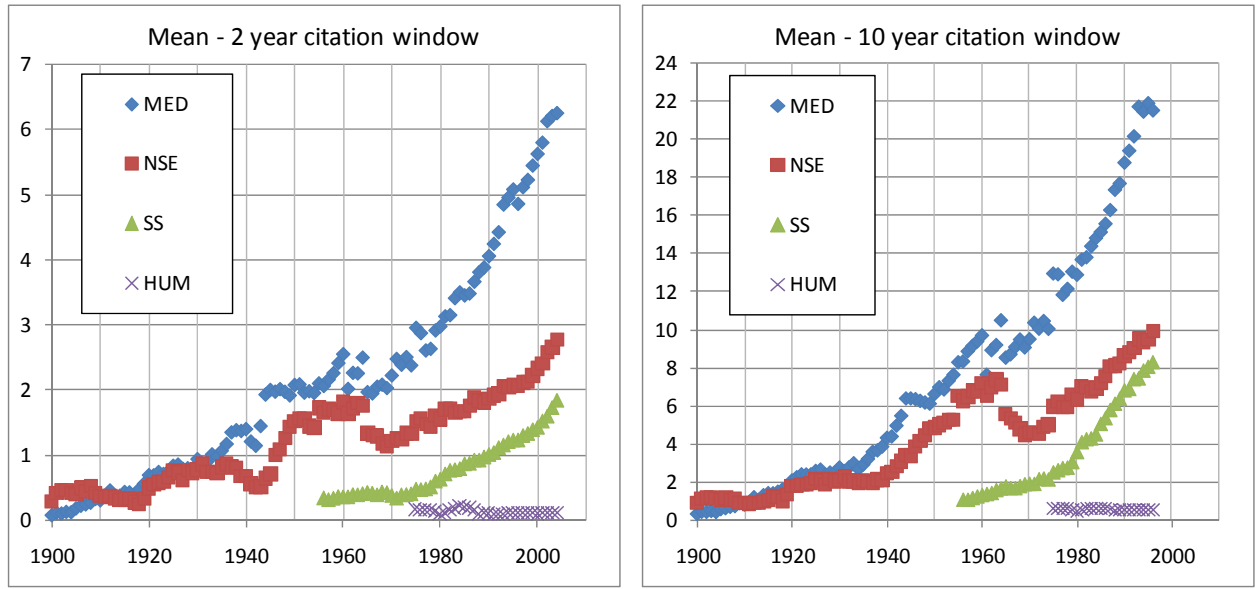

*Figure 1. Mean number of citations per paper, 2 years after publication (1900-2004) and 10 years after publication (1900-1996).*

Fig. 2 presents the data for 2-year and 10-year citation windows, divided into 6 classes that take into account the skewness of citation distributions: 0 citations, 1 citation, between 2 and 5 citations, between 6 and 10 citations, between 11 and 20 citations, and 21 citations or more. These data clearly show that, contrary to a widespread belief, uncitedness has generally *declined* for all disciplines but has remained more or less constant in HUM, although it must be noted that citation data for HUM are not reliable before 1988. These data confirm that in all other fields, science is increasingly drawing on the stock of published papers. Though not shown, data for a 5-year window show the same trends. In the particular case of HUM, the high level of uncitedness, as well as the relatively low number of citations received and their stability over time, are probably related to the fact that these disciplines tend to cite books more often than articles [24]. It is also readily apparent that in NSE, for instance, the level of uncitedness is as low today as it was between 1950 and 1965, albeit for entirely different reasons (see section 7).

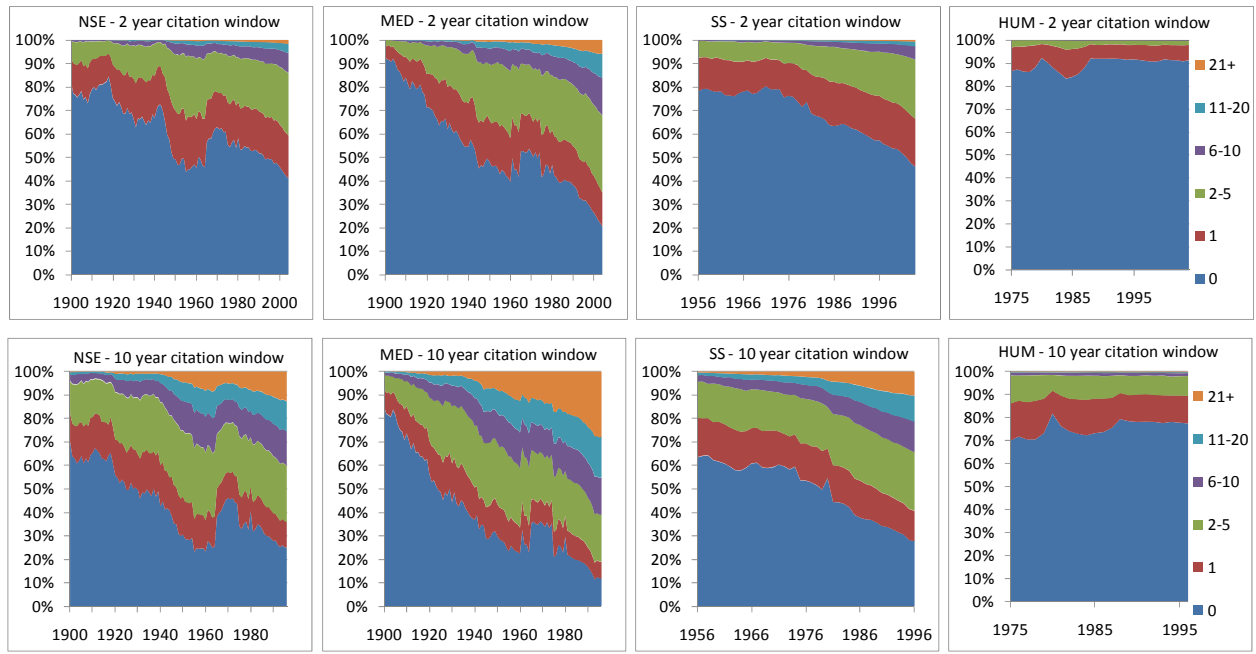

*Figure 2. Share of citations per paper 2 years after publication (1900-2004) and 10 years after publication (1900-1996)*

**Uncitedness on the decline**

The data presented so far in the paper show that for all fields but HUM, the number of citations received by papers underwent an overall increase during the 20th century, tempered by a few local fluctuations between 1900 and 1970. More significantly, the share of papers that remain uncited has decreased. In order to explain the changes in articles' uncitedness rates, the present section will provide a simple model that takes into account changes in the number of references per paper and in the growth of the world's scientific production [22]. Given that data for the full period is only available for NSE and MED, we will examine only these two fields in detail.

Previous studies [9, 25, 26] have shown that the first citation a paper receives likely occurs shortly after its publication, specifically according to a decay-type differential equation or a robust two or three-parameter function based on the articles' aging distribution. For our purposes, and based on the raw data from Fig. 2, we simply approximate the first citations as either "immediate" or "latent" (arising at least three years or so years after publication).

In the first case, our hypothesis is that the first citation depends simply on number of articles published in a given year ($N_A$) and the number of references available from the time of the paper's publication until two years thereafter (*i.e.* for a 2-year citation window) as $N_R$. The chance of citing a given paper within the pool of $N_A$ papers is thus a random selection process, expressed as simple Poisson distribution (since $N_R$ and $N_A$ are large). But only a fraction of references can be considered as randomly "choosing" recently-published papers, so the *effective* number of available references is $\beta_I N_R$, where $\beta_I$ is found empirically, via a least-squares fit, to be around 0.016 in the MED and 0.014 in the NSE — two very similar values to describe the citation practices in distinct scientific fields. The incidence of uncitedness, or the probability of getting zero immediate "successes", is thus expressed as

$$\Phi_I = e^{-\beta_I \frac{N_R}{N_A}} \qquad (1)$$

In Fig. 3, we plot the empirical data and Eq. (1), which shows very good agreement over the entire period, capturing many of the local trends which appear in the data. The deviations are more important in the case of MED than NSE, but only before 1915. In view of our approximations, this is simply due to the fact that citation practices seem to be less constant in time in MED than in NSE. In the latter, only the period between approximately 1960 and 1965 deviates systematically from the predictions of Eq. (1) – we will return to this particular case in Section 7. The very good agreement between data and Eq. (1) means that not only is uncitedness dominated by the ratio of references available to papers published (similar to the "Relative Publication Growth" developed in [27]), but also that the fraction of references used to determine first citations is approximately invariant over a period of 100 years. At the macroscopic level, further mechanisms that could affect whether a given paper will get cited or not seem to have relatively little impact.

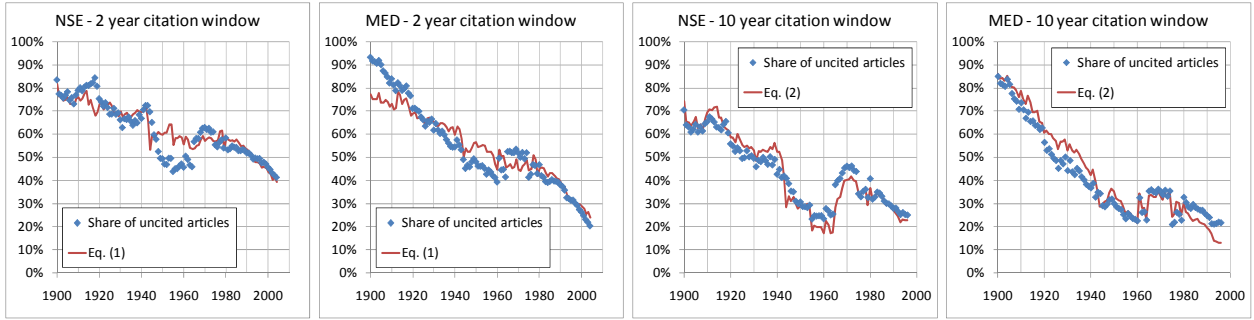

*Figure 3. Uncitedness data, compared with predictions, for 2-year and 10-year citation window. Predictions for the 10-year window are found by applying Eq. (2) to the 2-year data.*

The number of latent citations will inevitably depend once again on the ratio of references to remaining uncited articles, with even a smaller number of references from which to randomly choose old uncited articles. In a similar way, the chance of remaining uncited (after 10 years) can be expressed as the fraction of total ("latent" and "immediate") uncited papers:

$$P(n=0) = \Phi_T = \Phi_I e^{-\beta_L \frac{N_R}{N_A \Phi_I}} \qquad (2)$$

Fig.3 (10 year citation window) once again shows excellent agreement between raw data and the model above. Unsurprisingly, $\beta_L$ is much smaller than $\beta_I$ (by about one order of magnitude), since only a few references will have a "chance" of citing old, uncited papers. Furthermore, we have shown that capturing the main mechanisms responsible for uncitedness does not require complex mathematics or excessive parameterization; a simple, time-invariant approximation is robust enough to explain the bulk of over 100 years of uncitedness, implicitly averaged over many different areas of science.

At first glance, our results show that, in a given year, relatively high uncitedness is a consequence of relatively slow growth (*i.e.* not a sufficient amount of papers published in subsequent years) and a relatively stable number of references per paper. This is manifest in the case of the two World Wars where, as seen in Fig. 3 for instance, a lack of publication during the wars means higher uncitedness (during a two-year citation window) beginning around the two years preceding the hostilities. A sharp increase in the number of publications immediately after the wars means that papers published up to two years before will have a high chance of being cited. This does not, however, explain why uncitedness has been constantly declining for the past twenty years or so, since we know that the period of exponential growth in publishing has ended, and researchers are "forced" to cite older work. The answer lies in the fact that the average number of references per article has almost doubled between 1980 and 2004 in NSE and MED, having increased by the same factor in the previous 75 years [22]. In terms of uncitedness, this phenomenon has thus counterbalanced the slowing of scientific growth in recent years.

**The low and intermediate citation regime: universality from stretched-exponential functions**

Naturally, the probability of a paper having $n > 0$ citations cannot be described by a simple Poisson distribution without incorporating a second probability density function to describe the variable rates of citation [6-8], but we can use our results for uncitedness as a starting point to evaluate and elucidate some of the empirical and macroscopic citation models. We have found that not only is uncitedness sufficiently well-approximated by time-invariant and simple arguments, but that the $P(n = 0)$ case also fits the $P(n > 0)$ data perfectly well, *i.e.* they can be described by the same function(s) and are therefore derived from the same set of social mechanisms.

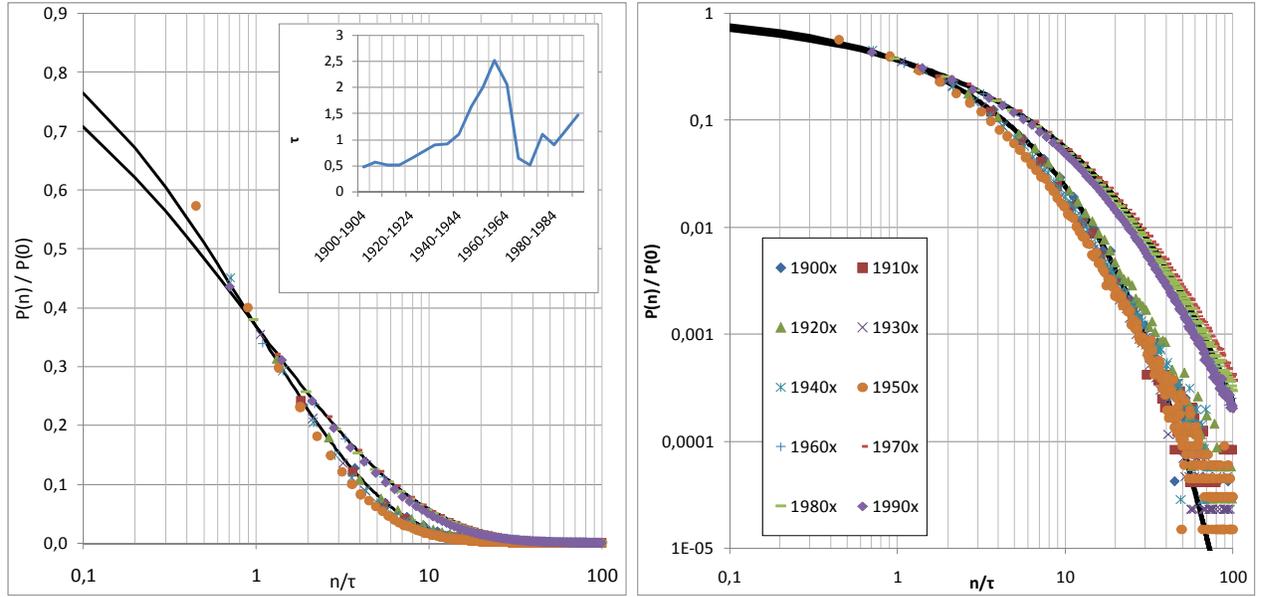

*Figure 4. Collapsed citation distribution over all decades, compared with two stretched-exponential fits using values of α: 0.47 (solid line) and 0.57 (dashed line), shown in semi-log and log-log form (for clarity). Note that the crossover point between the two master curves seems to occur around 1960. Inset (left): Evolution of τ, found via a linear least-squares fit, after re-arranging Eq. (3); note, once again, the peak around 1960.*

In order to perform a complete analysis of the citation distribution, we have focused our efforts on NSE data from a 10-year citation window, initially divided into 5-year periods (large enough for excellent statistics, but small enough to detect variations over time) from 1900 to 1994. Indeed, we have found that a stretched exponential function of the form,

$$P(n) = P(0) e^{-\left(\frac{n}{\tau}\right)^\alpha} \qquad (3)$$

fits the data very well ($R^2 > 0.98$) at worst for $n < 40$ (earlier in the century) and at best for $n < 200$ (during the 1990s). In the normalized distribution function (Fig. 4), $P(0)$ can be taken directly from uncitedness data or can be estimated with satisfactory precision using Eqs. (1) and (2). The two parameters $\tau$ and $\alpha$ are found empirically by rearranging the equation in terms of logarithms and performing a simple least-squares linear fit over the relevant regime (low and intermediate values of $n$). This function seems intuitively well-suited to the citation process, by either considering the distribution as the result of a series of stochastic (citation) processes [28] or, alternately, as an ensemble of papers gaining more citations at different exponential "rates" (analogous to a decay or relaxation time). We believe this stretched exponential regime to be the most crucial part of the citation "chain", since it includes virtually all papers and over 95% of citations, and thus captures the bulk of science. Perhaps too much attention has been paid to a small number of papers whose citations could be based on a power-law distribution arising, for instance, from a cumulative advantage process [2] or from the accumulation of "noise" in a set of stochastic processes [29].

Not only is the quality of the fit excellent during all periods, there is very little change in the values of $\tau$ except, once again, around 1960. When we rescale the number of citations by $\tau$ and normalize the distribution by $P(0)$ over the entire time period (this time by decade, for clarity), we note that there seem to be two regimes (Fig. 4), corresponding to different values of $\alpha$ with the crossover year around 1960 (see discussion below). The parameter $\tau$, on the other hand, indicates the characteristic "citation-scale" of the process, or the number of citations associated with the initial "decay" to higher-citedness. By rescaling, we isolate the *manner* in which the papers gain more citations (*how* these citation-scales are distributed). These are not rigorous interpretations of the data, but rather conceptual frameworks or analogies drawn from the physical sciences [30, 31]. Nevertheless, in light of the non-trivial results presented in Fig. 4, we believe this to be useful for comparing citation data at different times in history, and to shed light on how citation practices operate and how they have evolved over 100 years.

**A robust function to capture citation distribution tails and its application to "citation classics"**

There has also been growing evidence suggesting that the large-$n$ tails cannot be fit to the same function that dominates intermediate numbers of citations [11, 32]. We would prefer, however, to have one function able to fit data over the entire range of $n$. We find that the most robust fit (over all time periods) of the probability distribution function arises from the general, nonextensive statistical mechanics developed by Tsallis [32-34]. Without going into the details of Tsallis' approach (the reader is directed to [33-34]), we can say that, in our case, it would originate from the maximum entropy of a non-ergodic phase space formed by the stochastic "citation chain". It is expressed as,

$$P(n) = \frac{P(0)}{[1+(q-1)\lambda n]^{\frac{q}{q-1}}} \qquad (4).$$

In order to make sure we are getting a good fit at long time scales [35], we have used the cumulative probability density function $P_C(n)$ to generate a Zipf plot, and have accordingly transformed Tsallis' equation to,

$$P_C(n) = \frac{P(0)}{\lambda}\left[[1+(q-1)\lambda n]^{\frac{-1}{q-1}} - [1+(q-1)\lambda R]^{\frac{-1}{q-1}}\right] \qquad (5),$$

where $\lambda$ and $q$ are determined empirically from a non-linear fit (using the package provided in Microcal Origin) and $R$ denotes the maximum rank given to the citations or, equivalently, the maximum number of citations to one paper expected for a given data set. It need not be very precise and only becomes relevant for large $n$, since it brings a correction of at most $O(R^{-2})$ This approximate size limit on citations (equivalent to a cut-off of the power-law [29, 32] essentially introduces a finite-size effect to the system, generating the tails at large $n$ seen for many distributions in Fig. 5. The exponent $q$ varies between around 1.2 and 1.5, depending on the period studied, so there can be no universal power-law citation behavior, which seems consistent with varying citation practices.

We have thus shown that the Tsallis distribution is robust enough to fit the intermediate and high range of citation distributions over all time periods (without an excessive use of parameters or "extraneous" functions), and suggest further research into the details of specific citation mechanisms which can give rise to it. Contrary to Tsallis' claim, however, the function performs poorly at very small $n$ (see also Fig. 1 of [34]), where the bulk of citation occurs. When analytically calculating the average number of citations from Eq. (4), there appears to be a systematic overshoot, on the order of 5% to 20%, compared with the empirical value, which is nonetheless sufficiently accurate to observe all the main trends over 100 years (Fig. 1).

An alternative method was recently proposed [36], consisting in explicitly ranking papers in terms of their citations and applying a beta-like function, based on the "hierarchy" of a stochastic multiplicative process, over the entire range of ranks. We have found that this also fits citation data for most years extremely well, but a detailed discussion of the advantages of a rank-based approach (as opposed to standard probability distribution functions) remains beyond the scope of the present paper and is the subject of ongoing research.

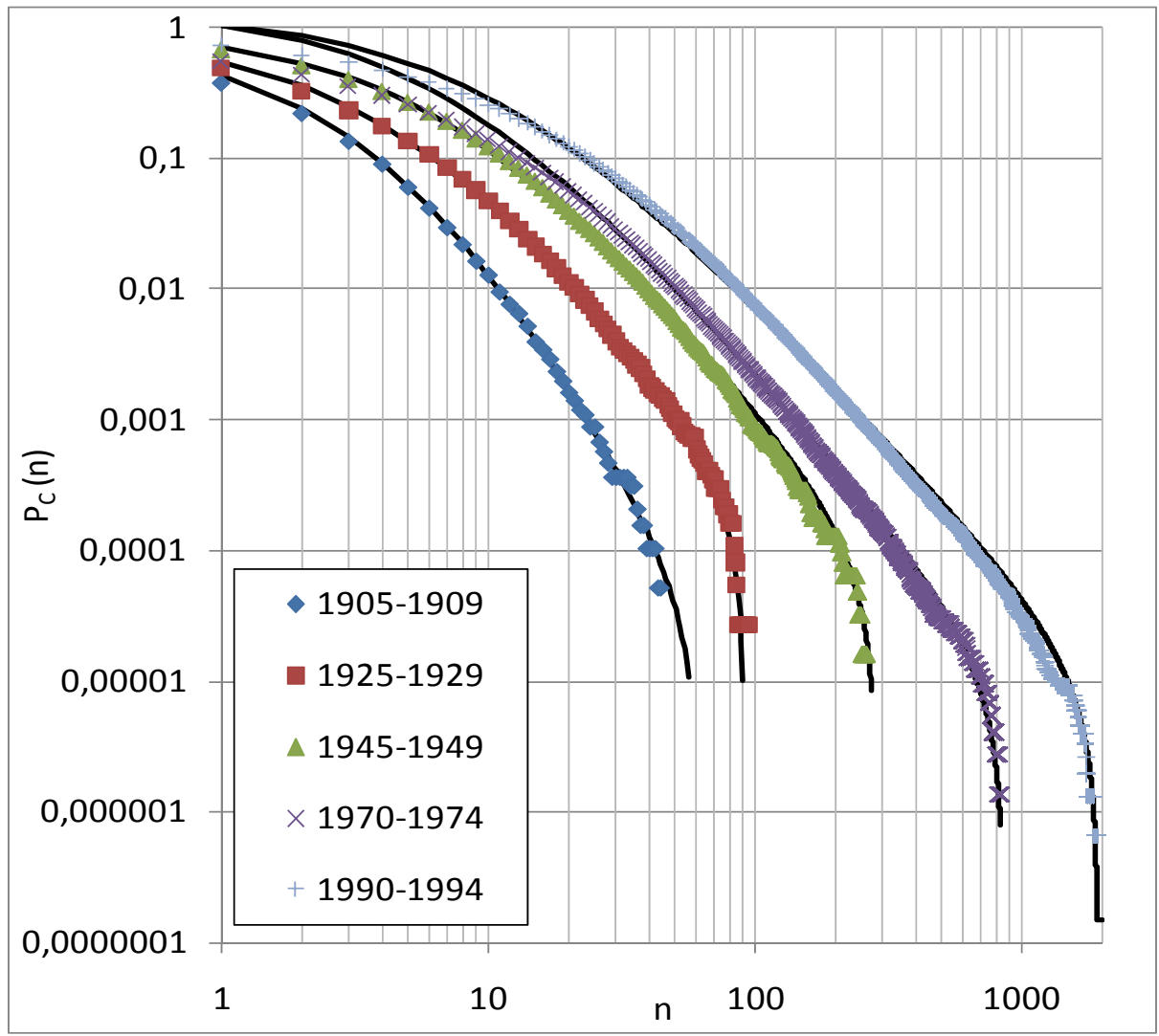

*Figure 5. Cumulative citation distributions and Tsallis fits from Eq. (5) at various times. It should be noted that the function is slightly less accurate for values of PC at very small n, which is difficult to see on a log-log plot.*

We propose to interpret the very large *n*-tails of the distribution as a useful and systematic way of determining what constitutes a citation "classic" [37, 38]. From Eq. (5), this is simply dependant on the distance from the maximum rank *R* and, once again, is equivalent to assigning some importance to the cutoffs of a power-law [29, 32]. Clearly, the number of citations which can intuitively (and mathematically) define a "classic" is not the same in 1920 as in 1980 (in MED, for instance, 20 citations has now become "average"!). The finite size of the pool of references means that this category of articles must drop off quickly (hence the tails). Our approach would thus empirically distinguish the "cream" from the rest of the crop, without requiring that the same number, or fraction, of articles every year achieve this status. As is shown in Fig. 5, certain periods, namely 1905-1909, display little evidence of these "classics" as distinct from the rest of the citation distribution. Overall, if one wishes to unambiguously define *any* given class of citations, applicable to different time periods or scientific fields, it is imperative to have a precise idea of the form of the *entire* citation distribution, hence the importance of our empirical demonstrations for sociologists or policymakers.

**Bibliometric data and the peak of postwar science**

The particularities in the citation data between around 1958 and 1965 in NSE (and, to a lesser extent, MED, although we have not examined this case in the same detail) are worthy of some special attention. First, certain characteristics, unmatched in the 20$^{th}$ century, can be ascribed to the evolution of science (but, seemingly, not of the social sciences – see Fig. 2) during this time. Furthermore, this exceptional case study serves to elucidate important elements of the trends we have observed over 100 years and to understand the significance (or limitations) of the models that can describe them.

Our approximations regarding the proportion of uncited papers – which perform surprisingly well during most of the 20th century – predict that, twenty years into the postwar "boom" of science, there would be an increase in uncitedness due to a *relative* and inevitable slowing-down of production. However, the model fails because the scientific community has proven itself to be particularly "efficient" during this period, meaning that even "marginal" authors were producing "citable" papers. Another way of looking at this is to consider certain periods of scientific activity to be more "compact" or cohesive than others, perhaps indicating that very few actors are working in the periphery of science. Take, for instance, the dominance of a few main scientific problems or extraordinarily fruitful avenues of research during this time: in physics, the development of the standard model and semiconductors; in biochemistry and biology, research on enzymes and proteins (thanks in part to rapid developments in organic chemistry methods), the discovery of the structure of DNA, and important progress in molecular biology (including molecular evolution and immunology).

In addition, Fig. 4 suggests that, around 1960, a sharp change in the distribution of citations occurred, separating two "steady-states" of citation practices. In other words, the impetus caused by the productivity boom forced a jump in the "citation-system", which then reconfigured to a state with a slightly lower stretching exponent $\alpha$. This generally indicates that the distribution of "characteristic citation-scales" (analogous to relaxation or decay times) changes to include larger citation-scales. Evidence for some type of transition appears in all our citation data, including simple indicators such as the average number of citations per paper during 2- and 10-year citation windows (Fig. 1).

## Conclusion

We have constructed a complete dataset of all citations to papers published between 1900 and 2006, which highlights the similarities and differences in citation practices in MED, NSE, SS and HUM, the latter being something of a special case since references to articles have always been relatively rare. An overall decrease in uncitedness (for 2, 5 and 10 years after publication) since 1900, as well as several local trends observed over 100 years can be satisfactorily explained by assuming that, each year, *the same fraction of references* contributes to randomly giving papers their first citation. Specifically, the trends observed during the wars, for instance, were due to changes in the number of *papers* published, while the recent decrease in uncitedness (contrary to what is often believed) is due to a higher number of *references per paper*. The relatively low level of uncitedness around the 1960s (compared with the mathematical predictions) might be explained by the efficiency or cohesiveness of the scientific community around this time.

Taking a holistic and macroscopic perspective on citation distributions (including uncitedness), we have insisted on the importance of the stretched-exponential function which dominates at small and intermediate *n* (the bulk of citations). We have used simple fits to extract data from this function and rescale the 100 years of citation data onto two master curves, finding a crossover point and exceptionally large values of $\tau$ once again around 1960. We have shown that, by using the *cumulative* distribution function, applying Tsallis' nonextensive statistical mechanics and considering a finite size limit for the number of citations, we can accurately model most of the citation distribution function (including the tails), at the expense of the very small *n* regime, and that such a fit is robust enough to be applied to citation data from all periods of the $20^{th}$ century. The tails of the Tsallis function, as deviations from a power-law, are furthermore proposed as a way to systematically and objectively identify highly-cited, "exceptional" papers.

The last example is indicative of how understanding the changes (or lack thereof) in citation processes (over 100 years) on a primarily phenomenological level has important implications for historians of science or those involved in research policy. Uncited papers and citation "classics", for instance, must be viewed as integral (yet distinguishable) parts of the "citation chain", be it on an empirical or theoretical level. Our results point to the possibility of a universal, more complete understanding of the mechanisms involved in the citation (or non-citation) of papers, either by considering how and when the stochastic "event" of selecting a paper for citation occurs, or what determines their citation *rank* within the field of science.